\documentclass[aps,twocolumn, floats,preprintnumbers,showpacs,prd]{revtex4}
\usepackage{graphicx}
\usepackage{url}
\usepackage{amsmath}
\usepackage{amsfonts}
\usepackage{amssymb}
\def\beq{\begin{equation}}
\def\eeq{\end{equation}}
\def\beqa{\begin{eqnarray}}
\def\eeqa{\end{eqnarray}}

\def\ltap{\ \raise.3ex\hbox{$<$\kern-.75em\lower1ex\hbox{$\sim$}}\ }
\def\gtap{\ \raise.3ex\hbox{$>$\kern-.75em\lower1ex\hbox{$\sim$}}\ }

\begin{document}
\preprint{FERMILAB-PUB-10-002-T}

\title{A Light Dilaton in Walking Gauge Theories}

\author{Thomas Appelquist$^{a}$ and Yang Bai$^b$
\\
\vspace{2mm}
${}^{a}$Department of Physics, Sloane Laboratory, Yale University, New Haven, CT, 06520, \\
${}^{b}$Theoretical Physics Department, Fermilab, Batavia, IL, 60510
}


\begin{abstract}
We analyze the existence of a dilaton in gauge theories with approximate infrared conformal symmetry. To the extent that these theories are governed in the infrared by an approximate fixed point (walking), the explicit breaking of the conformal symmetry at these scales is vanishingly small. If confinement and spontaneous chiral-symmetry breaking set in at some infrared scale, the resultant breaking of the approximate conformal symmetry can lead to the existence of a dilaton with mass parametrically small compared to the confinement scale, and potentially observable at the LHC.
\end{abstract}
\maketitle

{\it{\textbf{Introduction.}}}
The spontaneous breaking of an approximate continuous symmetry leads to the existence of
a light pseudo-Nambu-Goldstone-boson (PNGB). The light pion in QCD, for example, is
the PNGB associated with the spontaneous breaking of an approximate chiral symmetry. A long standing question is whether there can also exist a PNGB, the dilaton,
associated with the spontaneous breaking of an approximate dilatation, or scale symmetry of
some four dimensional gauge theories~\cite{Goldberger:2007zk}. This breaking could arise, for example, along with the
spontaneous breaking of chiral or other global symmetries of these theories.

Even with vanishing particle masses, however, the dilatation symmetry is present only classically,
broken explicitly by the renormalization scale entering at the quantum level. The divergence of the dilatation current ${\cal D}^{\mu}$ is proportional to the
$\beta$-function $\beta(\alpha)$. There could be an approximate dilatation symmetry if $\beta(\alpha)$ is, in some sense, small. If this approximate
symmetry is broken spontaneously, a light dilaton would emerge as a PNGB.
This notion was explored in the 1980's, with the smallness of the $\beta$ function being due to the particular field content of the theory, leading to
``slow" running of the coupling. The results were inconclusive ~\cite{Bardeen:1985sm, Yamawaki:1985zg, Holdom:1986ub,  Gusynin:1987em}.

Gauge theories with non-trivial infrared fixed points (IRFP's) provide a natural description of slow running \cite{Lane:1991qh, Appelquist:1996}. IRFP's are known to exist if the fermion content of the
theory is such as to make the fixed point weak, and therefore accessible in perturbation theory. For electroweak applications, though, the IRFP may have to be strong enough (the fermion number small enough) to trigger the spontaneous
breaking of  chiral symmetry. Walking theories are those in which the super-critical IRFP is close enough to criticality so that the scale of breaking is ``small". The IRFP then governs the theory for a range of momenta above the breaking scale.

Recent lattice studies
\cite{Appelquist:2007hu, Hietanen:2008mr} indicate that relatively strong IRFP's do appear in certain gauge theories.  Depending on the number of fermions, these fixed points can be sub-critical, or somewhat super-critical leading to walking behavior.
Here we examine the question of whether a walking gauge theory can lead to the appearance of a dilaton. Although the scale of chiral symmetry breaking is small relative to the scale characterizing the UV behavior, this small scale \emph{is} the physical confinement scale $\Lambda$. Therefore, one can safely conclude that the theory contains a dilaton only if it is parametrically light relative to \emph{this} scale.

 A four-dimensional field theory with a dilatation symmetry is also invariant under the larger group of conformal transformations. Since in the presence of an IRFP, the interacting theory flows to one with conformal symmetry, the fermion-number range that leads to this behavior while maintaining asymptotic freedom is referred to as the conformal window. The existence of either exact or approximate conformal symmetry has lead also to a study of these theories based on the AdS/CFT correspondence, where the dilaton is dual to a radion.

Our analysis of dilatation symmetry and its breaking is for a vector-like gauge theory with a critical fermion-number separating the conformal window from a phase with confinement and chiral symmetry breaking. But the chiral symmetry breaking is not essential. The discussion could also be framed in the context of, say, a chiral gauge theory where a critical fermion-number separates a conformal phase from one with confinement and massless composite fermion formation, but no chiral symmetry breaking~\cite{AppDuanSannino}.

{\it{\textbf{The Dilatation Current and its Divergence.}}}
We consider an $SU(N)$ gauge theory with $N_f$ massless Dirac fermions. We take them to be in the fundamental representation, although our discussion can be applied to other representations as well. The Lagrangian is
\beq
{\cal L}\,=\,-\frac{1}{4}\,G_{\mu\nu\,0}^aG^{\mu\nu}_{a\,0}\,+\,i\,
\sum_{j=1}^{N_f}\,\bar{\psi}^j_0\,(\partial_\mu\,\gamma^\mu\,+
\,i\,g_0\,t_a\,A^a_{\mu\,0}\,\gamma^\mu)\,\psi_{j\,0}\,,
\label{eq:NALagrangian}
\eeq
%
where the $0$-subscript denotes the unrenormalized  coupling constant and fields. It is corrected by higher-dimension, irrelevant operators. Since the dominant terms are dimension-4, there is an approximate, low-energy
dilatation symmetry at the classical level.

At the quantum level, even disregarding the irrelevant operators, the dilatation current ${\cal D}^\mu$, related to the symmetric energy-momentum tensor $\theta^{\mu\nu}$ by ${\cal D}^\mu = \theta^{\mu\nu}\,x_\nu$, has a non-vanishing divergence given by ~\cite{Collins:1976yq, Nielsen:1977sy}
\beqa
\partial_\mu\,{\cal D}^\mu\,=\,\theta^\mu_\mu \,=\, \frac{\beta{(\alpha)}}{4\,\alpha}\,G^a_{\mu\nu}\,G^{a\,\mu\nu} \,.
\label{eq:nADilatationAnomaly}
\eeqa
Here, $\alpha \equiv \alpha(\mu)$ is the renormalized gauge coupling defined at some scale $\mu$, $\beta(\alpha)$ is the renormalization-group (RG) $\beta$ function, and $G_{\mu\nu}^a$ is the renormalized field-strength tensor. It is given by $G_{\mu\nu}^a\equiv \partial_\mu A_\nu^a -\partial_\nu A_\mu^a- g (Z_1/Z_2)f_{abc}A^b_\mu A^c_\nu$, where $A_\mu^a$ is the renormalized gauge field related to $A^a_{\mu\,0}$
by the wave-function renormalization factor $Z_3^{1/2}$, $Z_2$ is the fermion wave-function renormalization factor, $Z_1$ is the fermion-gauge boson coupling renormalization factor. To render the connected matrix elements of the composite operator $G^a_{\mu\nu}\,G^{a\,\mu\nu}$ finite in perturbation theory, additional subtractions are necessary ~\cite{Collins:1976yq,Nielsen:1977sy}. The connected matrix elements of $\theta^\mu_\mu$ are then UV finite and independent of the RG scale $\mu$.

The PCAC relation in QCD is similar. With a quark-doublet field $\psi(x)$  and a common quark mass $m$, the divergence of the axial isospin current is given by
\beqa
\partial^\mu j^{5a}_\mu  = 2\,m\,j^{5a},
\label{eq:QCDpcac}
\eeqa
where
$j^{5a} =i\,\bar{\psi} \gamma^5\tau^a \psi$, with $\psi(x)$ the renormalized quark field, related to the bare field by $Z_{2}^{1/2}$. 
Here, $m\equiv m(\mu)$ is the renormalized quark mass with anomalous dimension $-\gamma(\mu)$, and the mass operator $j^{5a}$ has anomalous dimension $\gamma(\mu)$. Thus, Eq. (\ref{eq:QCDpcac}), the analog of Eq. (\ref{eq:nADilatationAnomaly}), is RG invariant. Since $m(\mu)$ is small, the axial current is approximately conserved, and chiral perturbation theory can be used to compute the mass of the pions.

{\it{\textbf{Infrared Fixed Points.}}}
 For an $SU(N)$ gauge theory with $N_f$ flavors in the fundamental representation, the $\beta$ function is ~\cite{Amsler:2008zzb}
\beqa
\beta{(\alpha)}&\equiv& \mu\,\frac{\partial \alpha}{\partial \mu}\equiv -\frac{b_0}{2\,\pi}\,\alpha^2\,-\,\frac{b_1}{4\,\pi^2}\,\alpha^3
\,+\,\cdots   \,,
\label{eq:twoloop}
\eeqa
where the two leading coefficients are scheme-independent with $b_0 = (11N-2N_f)/3$.
%
%
For $b_0 >0$ and small ($N_f$ just below $11 N/2$), a perturbative IRFP exists with strength $\alpha^* \approx - 2 \pi\,b_0/b_1$~\cite{Banks:1981nn}, and the theory becomes conformal in the infrared. The value of $\alpha^*$ increases as $N_f$ decreases, suggesting that it eventually exceeds a critical strength $\alpha_c$ for the spontaneous breaking of chiral symmetry and appearance of confinement.

The critical value $N^c_{f}$ at which this happens is not likely to be accessible in perturbation theory. Its determination through an analysis of the gauge coupling and associated $\beta$ function therefore requires some non-perturbative scheme for the definition of these quantities. Lattice simulations are currently being employed for this purpose for a variety of gauge groups and representation assignments for the fermions \cite{Appelquist:2007hu}.
Precise values for $N^c_f$ are not yet determined, but infrared conformal behavior is seen
for a range of $N_f$ values, while chiral symmetry breaking and confinement are seen at lower values. Preliminary lattice evidence is also consistent with the idea that the transition at $N_f = N^c_f$ is second-order or higher.

Walking sets in when  $N_f$ is close to but below $N^c_f$: $0< N^c_f - N_f \ll N^c_f$. (This can be viewed as requiring some fine tuning in theory space.) Adopting a non-perturbative scheme for the definition of $\alpha(\mu)$, we expect $0<\alpha^*-\alpha_c\ll\alpha_c\leq O(1)$. The scale of chiral symmetry breaking and the associated confinement scale are then of order the scale $\Lambda$ at which $\alpha(\mu)$ crosses $\alpha_c$. It is vanishingly small relative to the scale characterizing the perturbative UV behavior as $N_f\rightarrow N^c_{f}$, and $\alpha$  is governed by the IRFP for some range of scales above $\Lambda$.

In the neighborhood of an IRFP, either exact if $N_f > N^c_f$ or approximate in the case of walking, the simplest assumption is that the $\beta$ function has a linear zero. Lattice evidence (Refs~\cite{Appelquist:2007hu} and \cite{Hietanen:2008mr}) directly supports this assumption for the case $N_f > N_{f}^c$. We have
\beqa
\beta{(\alpha)} \,\simeq \, - \,s\,(\alpha^* - \alpha )
+ O\left((\alpha^* - \alpha )^2\right)
 \,,
\label{fig:linearbeta}
\eeqa
where $s > 0$ is the slope at $\alpha = \alpha^* $. In the walking case ($0< \alpha^* - \alpha_c \ll \alpha_c $), Eq.~(\ref{fig:linearbeta}) governs the evolution of $\alpha$ as $\mu \rightarrow \Lambda$ for a range of $\mu$ above $\Lambda$. There, the solution to the linearized RG equation takes the form
\beq
\alpha \, \simeq \, \alpha^* - (\alpha^* - \alpha_c ) \left(\frac{\mu}{\Lambda}\right)^s \,,
\eeq
 and the $\beta$ function is given approximately by
\beqa
\beta{(\alpha )} \,\simeq \, s\,(\alpha_c - \alpha^* ) \left(\frac{\mu}{\Lambda}\right)^s\,.
\label{fig:linearbetamu}
\eeqa

Since $0< s \leq O(1)$ and $0< \alpha^* - \alpha_c \ll \alpha_c$, the $\beta$ function is small
 for a range of $\mu$ above $\Lambda$, suggesting that conformal perturbation theory can be used to compute the mass of a PNGB dilaton in analogy to the pion of QCD. (It is also possible that the $\beta$-function zero is higher order as in the scenarios of Ref. \cite{Kaplan:2009kr}, with the corresponding $\beta$ function also small
 for a range of $\mu$ above $\Lambda$).

{\it{\textbf{PCAC and the Pion Mass.}}}
We first review briefly the derivation of the partially conserved axial current (PCAC) formula for the pion mass. Defining the pion decay constant via
\beq
\langle 0 | j^{5a}_\mu(x) | \pi^b(p)\rangle \,=\, -\,i\,f_\pi\,\delta^{ab}\,p_\mu\,e^{-ipx}\,,
\eeq
and using Eq.~(\ref{eq:QCDpcac}), one has
\beq
\langle 0 | j^{5a} (0) |  \pi^b(p=0) \rangle \,=\, -\,\frac{f_\pi\,m_\pi^2}{2\,m}\,\delta^{ab}\,.
\label{eq:PCAC1}
\eeq

With $Q^{5a}\equiv \int d^3x\,J^{5a}_0$, we have the commutation relation $[iQ^{5a}, J^{5b}]=\delta^{ab}\,\bar{\psi}\psi$. Using the local version of this relation and approximate current conservation,
\beq
i\partial^\mu\,\langle 0| {\rm T} j^{5a}_\mu(x) j^{5b}(0) |0\rangle \,\simeq\,\delta^4(x) \delta^{ab}\,\langle 0| \bar{\psi} \psi |0\rangle \,,
\eeq
where ``T" means time-ordered. Fourier transforming and assuming pion pole dominance, one obtains
\beq
\langle \pi^b (p=0) | j^{5a}(0) | 0\rangle \,\simeq\, \frac{\langle 0| \bar{\psi} \psi |0\rangle}{f_\pi}\,\delta^{ab}\,.
\label{eq:PCAC2}
\eeq
Comparing Eqs.~(\ref{eq:PCAC1}) and (\ref{eq:PCAC2}), the PCAC formula for the pion mass is given by
\beq
m^2_\pi \,\simeq\, -\, \frac{2\,m\,\langle 0| \bar{\psi} \psi |0\rangle}{f_\pi^2}\,.
\eeq
Since the operator $j^{5a}$ has anomalous dimension  $\gamma(\mu)$ equal and opposite to that of the mass $m \equiv m(\mu)$, the expression for $m_{\pi}^2$ is $\mu$ independent. With $\mu$ taken to be above a GeV or so, $m(\mu)$ is only a few MeV, and the above expression gives the leading contribution to the pion mass in chiral perturbation theory.

{\it{\textbf{PCDC and the Dilaton Mass.}}}
We can repeat the PCAC analysis to derive an expression for the dilaton mass based on a partially conserved dilatation current (PCDC).
Taking $N_f \lesssim N^c_f$ and assuming the existence of a dilaton state $|\sigma(p)\rangle$, we define the dilaton decay constant as~\cite{Crewther:1970gx}
\beqa
\langle 0 | \theta^{\mu\nu}(x) |\sigma(p)\rangle \equiv \frac{f_\sigma}{3}\,(p^\mu p^\nu- g^{\mu\nu}p^2)\,e^{-ipx}\,,
\label{eq:deriv}
\eeqa
where  $p^2=m_\sigma^2$ and $| 0 \rangle$ is the vacuum state corresponding to spontaneously broken chiral and dilatation symmetry. Taking the divergence of the corresponding matrix element of ${\cal D}^\mu = \theta^{\mu\nu}\,x_\nu$ and using Eq.~(\ref{eq:nADilatationAnomaly}),  we have
\beqa
\partial_\mu \langle 0 | {\cal D}^\mu(x) |\sigma(p)\rangle = \langle 0 | \theta^\mu_\mu(x) |\sigma(p)\rangle = -\,f_\sigma\,m_\sigma^2\,e^{-ipx}.
\label{eq:relation1}
\eeqa

To proceed, it is natural to think that $G^a_{\mu\nu}G^{a\mu\nu}(x)$ will be the analogue of $\bar{\psi}\psi(x)$ in the PCAC discussion. But although the connected matrix elements of $G^a_{\mu\nu}G^{a\mu\nu}(x)$ such as  $\langle 0 |G^a_{\mu\nu}G^{a\mu\nu}(x)|\sigma(p)\rangle$  are finite, this is not true of its vacuum expectation value, which plays the defining role in spontaneous dilatation-symmetry breaking~\footnote{This is not a issue for the vacuum expectation value of $\bar{\psi}\psi(x)$ because of the chiral symmetry.}. It is quartically cutoff dependent in perturbation theory.

 We remove this piece by a subtraction procedure. We have defined the running coupling and the renormalized field strength $G^a_{\mu\nu}$ at some scale $\mu$ (with $\theta^{\mu}_{\mu}$ being $\mu$-independent). From here on, for simplicity, we take $\mu = \lambda \Lambda$ with $\lambda \geq 1 $,  satisfying $\alpha(\lambda\Lambda)\approx \alpha(\Lambda)$. We then define
\beqa
 [\theta^{\mu}_{\mu}]_{\Lambda}
& \equiv& \frac{\beta{(\alpha)}}{4\,\alpha} [G^a_{\mu\nu}G^{a\mu\nu}]_{\Lambda} \nonumber \\
 &\equiv& \frac{\beta{(\alpha)}}{4\,\alpha}\,G^a_{\mu\nu}G^{a\mu\nu}
 - \langle 0| \frac{\beta{(\alpha)}}{4\,\alpha}\,G^2|0 \rangle_{p\geq \lambda \Lambda}\,,
 \label{eq:subtract}
 \eeqa
where $\langle 0|\beta(\alpha)/4\alpha\,G^2|0 \rangle_{p\geq \lambda\Lambda}$ is the vacuum value computed by including momentum components only above  $\lambda\Lambda$  This subtraction should be implementable in a non-perturbative framework such as a lattice-based simulation, as well as in perturbation theory.

The subtraction leaves $\langle 0|[\theta^{\mu}_{\mu}]_{\Lambda}| 0 \rangle$ UV-cutoff independent and free of dependence on the high-energy RG scale associated with perturbative running which breaks the dilatation symmetry explicitly. This definition is useful only in the case of walking, where a hierarchy develops between $\Lambda$ and the high-energy scale. The quantity $\langle 0|[G^a_{\mu\nu}G^{a\mu\nu}]_{\Lambda}|0 \rangle$ serves as an order parameter for spontaneous dilatation-symmetry breaking, vanishing continuously as $N_f \rightarrow N^c_{f}$ from below ($\Lambda \rightarrow 0$). It has been defined in analogy to subtraction schemes defining the ``gluon condensate" in QCD ~\cite{Banks:1981zf}, but in a way appropriate for the study of spontaneous dilatation symmetry breaking.

To proceed as in the PCAC discussion, it is useful to define a correspondingly subtracted energy-momentum tensor $[\theta^{\mu \nu}(x)]_{\Lambda}=\theta^{\mu \nu}(x)-(g^{\mu\nu}/4)\langle 0| (\beta{(\alpha)} /4\alpha)G^2|0 \rangle_{p\geq \lambda \Lambda}$, and subtracted dilatation current $[{\cal D}^\mu (x)]_\Lambda =  x_{\nu}\,[\theta^{\mu \nu}(x)]_{\Lambda}$, whose divergence is small in all matrix elements including its vacuum value.
The corresponding charge
$[Q]_{\Lambda}=\int dx^3\, [{\cal D}^{\mu=0}(x)]_{\Lambda}$ is the generator of dilatation transformations. Since the underlying theory is approximately conformal at momentum scales of order  $\Lambda$ and  the operator $[\theta^{\mu}_{\mu}]_{\Lambda}$  has vanishing anomalous dimension, we have $[ i [Q]_{\Lambda}, [\theta^{\mu}_{\mu}]_{\Lambda}] \simeq \,4[\theta^{\mu}_{\mu}]_{\Lambda}$.

Making use of the local version of this commutation relation and using $\partial_\mu\,[{\cal D}^\mu(x)]_\Lambda \simeq 0$, we have
\beqa
 i\partial_\mu \langle 0| \mbox{T}\,[{\cal D}^\mu (x)]_{\Lambda} [\theta^{\mu}_{\mu}(0)]_{\Lambda}| 0 \rangle
 \simeq 4\,\delta^4(x)\,\langle 0| [\theta^{\mu}_{\mu}]_{\Lambda}| 0 \rangle .
\label{eq:vacvalue}
\eeqa
The contribution of the vacuum intermediate state on the LHS is small (and, in fact, is  exactly canceled by the corresponding contribution to the neglected term, $i\langle 0|{\rm T}[\theta^{\mu}_{\mu}(x)]_{\Lambda} [\theta^{\mu}_{\mu}(0)]_{\Lambda}|0\rangle$, on the RHS). Fourier transforming and assuming dilaton-state dominance of the remaining piece on the LHS, and using Eq.~(\ref{eq:deriv}), we obtain
\beqa
\langle \sigma(p=0) | [\theta^\mu_\mu(0)]_{\Lambda} |0\rangle \,\simeq\, \frac{4}{\,f_\sigma}\,\langle 0| [\theta^{\mu}_{\mu}]_{\Lambda}| 0 \rangle\,.
\label{eq:relation2}
\eeqa
Comparing Eq. (\ref{eq:relation2}) with  Eq. (\ref{eq:relation1}) (where
$\theta^{\mu}_{\mu}$ can be replaced with $[\theta^{\mu}_{\mu}]_{\Lambda}$), we arrive at the PCDC formula for the dilaton mass
\beqa
m_\sigma^2 \, \simeq \, -\,\frac{4}{\,f^2_\sigma}\,\langle 0| [\theta^{\mu}_{\mu}]_{\Lambda}| 0 \rangle\,.
\label{eq:dilatonmass}
\eeqa
%

  For an order of magnitude estimate, we take $f_\sigma \simeq \Lambda$. (In QCD $f_\pi \simeq 0.3 \Lambda$ ). From the definition Eq.~(\ref{eq:subtract}), we then expect $\langle 0|[G^a_{\mu\nu}G^{a\mu\nu}]_{\Lambda}|0 \rangle \simeq  \Lambda^4$, since there remains no larger momentum scale after the subtraction. Making the simplest assumption that the IRFP is described by a linear zero of the $\beta$ function, we can use Eq. (\ref{fig:linearbetamu}) to arrive at
\beqa
m_\sigma^2 &\simeq &  \frac{s\,(\alpha^* - \alpha_c)}{\alpha_c}\,\Lambda^2
\, \simeq \,  \frac{(N^c_{f} - N_{f})}{N^c_{f}}\,\Lambda^2
\,,
\label{eq:dilatonmass2}
\eeqa
where we have used the fact that for $N_f$ close to $N^c_{f}$, $(\alpha^* - \alpha_c)/\alpha_c \simeq O(1)(N^c_{f} - N_f)/N^c_{f}$. While this analysis has invoked scheme-dependent quantities such as $\alpha$ and $\alpha_c$, the final estimate, in terms of the physical confinement scale $\Lambda$, must be scheme independent.

The parametric smallness of $m_{\sigma}$ relative to the confinement and chirality-breaking scale
$\Lambda$ (the electroweak breaking scale in a technicolor context) is due to the parametric smallness of $\langle 0| [\theta^{\mu}_{\mu}]_{\Lambda}| 0 \rangle  = (\beta(\alpha)/4\,\alpha) \langle 0|[G^a_{\mu\nu}G^{a\mu\nu}]_{\Lambda}| 0 \rangle $. It is defined at a scale $\lambda \Lambda$ with $\lambda \geq 1$, and does not contain important contributions from scales below $\Lambda$. In the limit $N_f \rightarrow N^c_{f}$, since $\Lambda \rightarrow 0$, the order parameter $\langle0|[G^a_{\mu\nu}G^{a\mu\nu}]_{\Lambda}|0 \rangle$ vanishes, but the dilaton mass vanishes more rapidly. By contrast, in the PCAC case, $\langle 0| \bar{\psi} \psi |0\rangle$ remains finite as $m \rightarrow 0$.

Expressions similar to Eq. (\ref{eq:dilatonmass}) may be found in the literature \cite{Gusynin:1987em, Haba:2010hu}, but not employed in the present framework of approximate infrared  dilatation symmetry and not indicating parametric smallness of $m_{\sigma}$. The physical picture here is that the dilaton is formed at scales $\gtrsim  \Lambda$. The vector resonances, baryons, glueball states, etc. are also expected to be formed at these scales, as are the chiral NGB's. They and the PNGB dilaton are the only lighter states. The dilaton, with its vacuum quantum numbers, is an admixture of gluon and fermion constituents, but parametrically lighter than other $0^{++}$ states.

It will be important to explore and confirm these ideas in a more dynamical framework, for example through a study of the vacuum energy functional. This was attempted in Ref. \cite{Holdom:1986ub}, but using a quasi-perturbative approach with uncontrolled truncations and questions of gauge dependence. The only way we know to do this reliably is through lattice simulations. The gauge-boson condensate
$\langle 0|G^a_{\mu\nu}G^{a\mu\nu}|0 \rangle$, both unsubtracted and subtracted, will play a central role
here.  More directly, the existence of a parametrically light dilaton in a walking theory can be studied by the simulation of correlation functions involving operators with vacuum ($0^{++}$) quantum numbers. The spectral analysis should exhibit a parametrically light state as well as, say, glueball states with $O(\Lambda)$ masses.


{\it{\textbf{Higher-Dimension Operators.}}}
A gauge theory is naturally corrected by higher-dimension (irrelevant) operators entering at some ultraviolet scale. In technicolor, these are the effective four-fermion  interactions responsible for quark and lepton masses, and also for lifting the masses of some chiral NGB's (those not eaten by the $W$ and $Z$).
Since these operators explicitly break dilatation symmetry, they also contribute to the mass of the dilaton. An application of Dashen's formula \cite{Dashen:1969eg} gives
$\Delta m_\sigma^2 = \langle 0|[Q,[Q,{\cal H}]] |0 \rangle\ / f^2_\sigma$, where ${\cal H}$ is a higher-dimension term in the Hamiltonian. The contribution to the dilaton mass from such terms is suppressed providing either that the anomalous dimension of ${\cal H}$ is small enough at the UV scale to maintain its dimension above $4$ or that there is a small overall coefficient.

{\it{\textbf{Conclusion and Discussion.}}}
We have argued that a parametrically light dilaton is a natural feature of a walking gauge theory, one characterized by an approximate infrared fixed point somewhat supercritical for the spontaneous breaking of the chiral symmetries. There is no such state in a QCD-like theory. The dilaton is narrow, decaying into the chiral NGB's in the model of this paper. With standard-model fields included, it can potentially decay into longitudinal gauge bosons, chiral PNGB's, quarks and leptons. The couplings of the dilaton field to the standard-model fields can be obtained from the standard-model Higgs couplings by replacing $v_{\rm EW}$ by $f_\sigma$. Its LHC phenomenology, dark matter consequences, and flavor constraints have been studied recently in  Refs. \cite{Goldberger:2007zk, Fan:2008jk, Bai:2009ms, Vecchi:2010gj,Elander:2009pk}.

\vspace*{0.06in}
We thank Bill Bardeen, George Fleming, Walter Goldberger, Ethan Neil, Witek Skiba, and Rohana Wijewardhana for helpful discussions. We also thank the Aspen Center for Physics where this work was initiated. This work (TA) was supported partially by DOE grant DE-FG02-92ER-40704. Fermilab is operated by Fermi Research Alliance under contract no. DE-AC02-07CH11359 with the United States Department of Energy.

\end{document}